\pdfoutput=1

\documentclass[11pt,twoside,a4paper,cmspaper,final,collab]{cms-tdr}
\def\svnVersion{446759P}\def\cmsCernNoTag{CERN-EP-2017-207}\def\cmsCernDate{\today}\def\cmsMessage{Published in Physical Review Letters as \href{http://dx.doi.org/10.1103/PhysRevLett.120.071802}{\doi{10.1103/PhysRevLett.120.071802.}}}

\begin{document}\cmsNoteHeader{HIG-17-010}

\hyphenation{had-ron-i-za-tion}
\hyphenation{cal-or-i-me-ter}
\hyphenation{de-vices}
\RCS$Revision: 440224 $
\RCS$HeadURL: svn+ssh://svn.cern.ch/reps/tdr2/papers/HIG-17-010/trunk/HIG-17-010.tex $
\RCS$Id: HIG-17-010.tex 440224 2017-12-26 14:51:51Z alverson $
\newlength\cmsFigWidth
\newlength\cmsTabWidth
\ifthenelse{\boolean{cms@external}}{\setlength\cmsFigWidth{0.96\columnwidth}}{\setlength\cmsFigWidth{0.48\textwidth}}
\ifthenelse{\boolean{cms@external}}{\setlength\cmsTabWidth{0.96\columnwidth}}{\setlength\cmsTabWidth{0.6\textwidth}}
\ifthenelse{\boolean{cms@external}}{\providecommand{\cmsLeft}{top\xspace}}{\providecommand{\cmsLeft}{left\xspace}}
\ifthenelse{\boolean{cms@external}}{\providecommand{\cmsRight}{bottom\xspace}}{\providecommand{\cmsRight}{right\xspace}}
\newcommand{\PV}{\ensuremath{\cmsSymbolFace{V}}\xspace}
\ifthenelse{\boolean{cms@external}}{\providecommand{\NA}{\ensuremath{\cdots}\xspace}}{\providecommand{\NA}{\ensuremath{\text{---}}\xspace}}
\ifthenelse{\boolean{cms@external}}
{\providecommand{\suppMaterial}{the supplemental material~\cite{supp}, which includes Ref.~\cite{Bahr:2008pv}}}
{\providecommand{\suppMaterial}{Appendix~\ref{app:suppMat}}}

\newcommand{\mH}{\ensuremath{m_{\PH}}\xspace}
\newcommand{\mSD}{\ensuremath{m_{\mathrm{SD}}}\xspace}
\newcommand{\nddt}{\ensuremath{N_{2}^{1\mathrm{,DDT}}}\xspace}
\newcommand{\muVal}{\ensuremath{2.3}\xspace}
\newcommand{\muStatErrLo}{\ensuremath{1.5}\xspace}
\newcommand{\muSystErrLo}{\ensuremath{0.4}\xspace}
\newcommand{\muErrLo}{\ensuremath{1.6}\xspace}
\newcommand{\muErrHi}{\ensuremath{1.8}\xspace}
\newcommand{\muSystErrHi}{\ensuremath{1.0}\xspace}
\newcommand{\muObsLimit}{\ensuremath{5.8}\xspace}
\newcommand{\muExpLimit}{\ensuremath{3.3}\xspace}
\newcommand{\muNoObsLimit}{\ensuremath{7.2}\xspace}
\newcommand{\muNoExpLimit}{\ensuremath{4.1}\xspace}
\newcommand{\muNoVal}{\ensuremath{3.2}\xspace}
\newcommand{\muNoErrLo}{\ensuremath{2.0}\xspace}
\newcommand{\muNoErrHi}{\ensuremath{2.2}\xspace}
\newcommand{\muNoObsSig}{\ensuremath{1.6}\xspace}
\newcommand{\muNoExpSig}{\ensuremath{0.5}\xspace}
\newcommand{\xsecThyVal}{\ensuremath{31.7}\xspace}
\newcommand{\xsecThyErr}{\ensuremath{9.5}\xspace}
\newcommand{\xsecVal}{\ensuremath{74}\xspace}
\newcommand{\xsecStatErrLo}{\ensuremath{48}\xspace}
\newcommand{\xsecSystErrLo}{\ensuremath{10}\xspace}
\newcommand{\xsecSystErrHi}{\ensuremath{17}\xspace}
\newcommand{\xsecZThyVal}{\ensuremath{1.09}\xspace}
\newcommand{\xsecZThyErr}{\ensuremath{0.11}\xspace}
\newcommand{\xsecZVal}{\ensuremath{0.85}\xspace}
\newcommand{\xsecZStatErrLo}{\ensuremath{0.16}\xspace}
\newcommand{\xsecZSystErrLo}{\ensuremath{0.14}\xspace}
\newcommand{\xsecZSystErrHi}{\ensuremath{0.20}\xspace}
\newcommand{\muObsSig}{\ensuremath{1.5}\xspace}
\newcommand{\muExpSig}{\ensuremath{0.7}\xspace}
\newcommand{\muZVal}{\ensuremath{0.78}\xspace}
\newcommand{\muZStatErrLo}{\ensuremath{0.14}\xspace}
\newcommand{\muZSystErrLo}{\ensuremath{0.13}\xspace}
\newcommand{\muZErrLo}{\ensuremath{0.19}\xspace}
\newcommand{\muZErrHi}{\ensuremath{0.23}\xspace}
\newcommand{\muZSystErrHi}{\ensuremath{0.19}\xspace}
\newcommand{\muZObsSig}{\ensuremath{5.1}\xspace}
\newcommand{\muZExpSig}{\ensuremath{5.8}\xspace}

\cmsNoteHeader{HIG-17-010}
\title{Inclusive search for a highly boosted Higgs boson decaying to a bottom quark-antiquark pair}

\date{\today}

\abstract{
   An inclusive search for the standard model Higgs boson ($\PH$) produced with large transverse momentum ($\pt$) and decaying to a bottom quark-antiquark pair ($\bbbar$) is performed using a data set of $\Pp\Pp$ collisions at $\sqrt{s}=13\TeV$ collected with the CMS experiment at the LHC. The data sample corresponds to an integrated luminosity of 35.9\fbinv. A highly Lorentz-boosted Higgs boson decaying to $\bbbar$ is reconstructed as a single, large radius jet and is identified using jet substructure and dedicated $\PQb$ tagging techniques. The method is validated with $\PZ\to\bbbar$ decays. The $\PZ\to\bbbar$ process is observed for the first time in the single-jet topology with a local significance of \muZObsSig standard deviations ($\muZExpSig$ expected). For a Higgs boson mass of 125\GeV, an excess of events above the expected background is observed (expected) with a local significance of \muObsSig ($\muExpSig$) standard deviations. The measured cross section times branching fraction for production via gluon fusion of $\PH \rightarrow \bbbar$ with reconstructed $\pt>450\GeV$ and in the pseudorapidity range $-2.5 < \eta  < 2.5$ is $\xsecVal\pm\xsecStatErrLo\stat_{-\xsecSystErrLo}^{+\xsecSystErrHi}\syst\unit{fb}$, which is consistent within uncertainties with the standard model prediction.
}

\hypersetup{%
pdfauthor={CMS Collaboration},%
pdftitle={Inclusive search for a highly boosted Higgs boson decaying to a bottom quark-antiquark pair},
pdfsubject={CMS},%
pdfkeywords={CMS, boosted Higgs, bottom quark-antiquark decay, gluon fusion}}

\maketitle

In the standard model (SM)~\cite{Salam:1968rm,Glashow:1961tr,Weinberg:1967tq}, the Brout-Englert-Higgs mechanism ~\cite{PhysRevLett.13.321,Higgs:1964ia,PhysRevLett.13.508,PhysRev.145.1156,PhysRevLett.13.585} is responsible for electroweak symmetry breaking and the mass of elementary particles. Although a Higgs boson ($\PH$) was discovered~\cite{:2012gk,:2012gu,Chatrchyan:2013lba}, the LHC data sets of $\Pp\Pp$ collisions at $\sqrt{s} = 7$ and 8\TeV were not sufficient to establish the coupling to bottom quarks~\cite{CMS:2015kwa}, despite the 58.1\% expected branching fraction of the Higgs boson to bottom quark-antiquark ($\bbbar$) pairs~\cite{deFlorian:2016spz}. The most sensitive method to search for $\PH\to\bbbar$ decays at a hadron collider is to use events in which the Higgs boson is produced in association with a $\PW$ or $\PZ$ boson ($\PV\PH$) decaying to leptons, and recoiling with a large transverse momentum (\pt)~\cite{Butterworth:2008iy}, in order to suppress the overwhelming irreducible background from quantum chromodynamics (QCD) multijet production of $\PQb$ quarks. Because of this background, an observation of $\PH(\bbbar)$ decays in the gluon fusion production mode (ggF) was considered impossible. This Letter presents the first inclusive search for $\PH\to\bbbar$, where the Higgs boson is produced with high-$\pt$. Measurements of high-$\pt$ $\PH(\bbbar)$ decays may resolve the loop induced and tree-level contributions to the ggF process~\cite{Grojean:2013nya} and provide an alternative approach to study the top quark Yukawa coupling in addition to the $\ttbar\PH$ process.

The results reported in this Letter are based on a data set of $\Pp\Pp$ collisions at $\sqrt{s} = 13\TeV$ collected with the CMS detector at the LHC in 2016 and corresponding to an integrated luminosity of 35.9\fbinv.
The main experimental difficulties for this search originate from the large cross section for background multijet events at low jet mass and the restrictive
trigger requirements needed to reduce the data recording rate.
Therefore, we require events to have a high-$\pt$ Higgs boson candidate and define six $\pt$ categories from 450\GeV to 1\TeV with variable width from 50 to 200\GeV.
Combinatorial backgrounds are reduced by requiring the Higgs boson's decay products to be clustered in a single jet~\cite{Butterworth:2008iy}.
The jet is required to have two-prong substructure and $\PQb$ tagging properties consistent with the $\PH(\bbbar)$ signal. The nontrivial jet mass shape is difficult to model parametrically. For this reason, the dominant background from SM QCD multijet production is estimated in data by inverting the $\PQb$ tagging requirement, which is, by design, decorrelated from jet mass and $\pt$.
A simultaneous fit to the distributions of the jet mass in all categories is performed in
the range 40 to 201\GeV to extract the inclusive $\PH(\bbbar)$ and $\PZ(\bbbar)$ production cross sections and to determine the normalizations and shapes of the jet mass distributions for the backgrounds.

A detailed description of the CMS detector, together with a definition of the coordinate system and the relevant kinematic variables, can be found in Ref.~\cite{Chatrchyan:2008zzk}.
The central feature of the CMS apparatus is a superconducting solenoid of 6\unit{m} internal diameter, providing a magnetic field of 3.8\unit{T}. A silicon pixel and strip tracker, a lead tungstate crystal electromagnetic
calorimeter, and a brass and scintillator hadron calorimeter, each composed of a barrel and two endcap sections, reside within the solenoid. Forward
calorimeters extend the pseudorapidity ($\eta$)~\cite{Chatrchyan:2008zzk}
coverage provided by the barrel and endcap detectors. Muons are detected in gas-ionization chambers embedded in the steel flux-return yoke outside the solenoid.

Simulated samples of signal and background events are produced using various Monte Carlo (MC) event generators, with the CMS detector response modeled using the \GEANTfour~\cite{GEANT4} program.
The \MGvATNLO~2.3.3~\cite{Alwall:2014hca} generator is used for the diboson, $\PW+$jets, $\cPZ+$jets, QCD multijet samples at leading order (LO) accuracy with matching~\cite{Alwall:2007fs} between jets from the matrix element calculation and the parton shower description,
while \POWHEG 2.0~\cite{Nason:2004rx,Frixione:2007vw,Alioli:2010xd} at next-to-leading order (NLO) precision is used to model the $\ttbar$ and single-top processes. For parton showering and hadronization the \POWHEG and \MGvATNLO samples are interfaced with \PYTHIA 8.212~\cite{Sjostrand:2014zea}.
The \PYTHIA parameters for the underlying event description are set to the
CUETP8M1 tune~\cite{Khachatryan:2015pea}.
The production cross sections for the diboson samples are calculated to next-to-next-to-leading-order (NNLO) accuracy with the \textsc{mcfm}~7.0 program~\cite{Campbell:2010ff}. The cross section for top quark pair production is computed with \textsc{Top++} 2.0~\cite{Czakon:2013goa} at NNLO. The cross sections for $\PW+$jets and $\cPZ+$jets samples include higher-order QCD and electroweak (EW)
corrections and improve modeling of high-$\pt$ $\PW$ and $\PZ$ bosons events~\cite{Kallweit:2014xda,Kallweit:2015dum,Kallweit:2015fta,Lindert:2017olm}.
The parton distribution function (PDF) set {NNPDF3.0}~\cite{Ball:2014uwa} is used to produce all simulated samples, with the accuracy (LO or NLO) corresponding to that of the generator used.
The Higgs boson signal samples are produced using the \POWHEG event generator with $\mH = 125\GeV$. For the ggF production mode, the \POWHEG generated sample with up to one extra jet in matrix element calculations is normalized to the inclusive cross section at next-to-next-to-next-to-leading order (N$^{3}$LO) accuracy~\cite{deFlorian:2012mx,Grazzini:2013mca,Bagnaschi2012,Bagnaschi:2015qta}. The resulting Higgs boson $\pt$ spectrum neglects the effects of the finite top quark mass~\cite{Baur:1989cm} and associated higher-order QCD corrections~\cite{Boughezal:2013uia,Boughezal:2015aha,Boughezal:2016wmq,Chen:2016vqn}, which are expected to be large for $\pt$ greater than approximately twice the mass of the top quark~\cite{Baur:1989cm}. A $\pt$-dependent correction has been derived to account for both of these effects.
The \POWHEG ggF $\pt$ spectrum is reweighted to the 0--2~jet CKKW-L~\cite{Catani:2001cc,Lonnblad:2001iq,Czakon:2013goa} merged LO ggF process incorporating the finite top quark mass ($m_{\cPqt}$)~\cite{Hirschi:2015iia,Buschmann:2014sia,deFlorian:2016spz,Greiner:2015jha}. This spectrum is then corrected by the approximate NLO to LO ratio, obtained by expanding in powers of $1/m_{\cPqt}^2$ up to $1/m_{\cPqt}^4$, and the effective NNLO to NLO ratio~\cite{Neumann:2016dny,Greiner:2016awe} in the infinite top quark mass approximation.
The overall \pt-dependent correction to the initial N$^{3}$LO \POWHEG ggF spectrum is found to be $1.27\pm0.38$, resulting in a ggF cross section times $\PH(\bbbar)$ branching fraction of $\xsecThyVal\pm\xsecThyErr\unit{fb}$ for reconstructed Higgs boson $\pt > 450 \GeV$ and $\abs{\eta}<2.5$. An  uncertainty of 30\% to the overall correction is estimated from the comparison of different predictions obtained by using: (i) a merging scale of 100 instead of 20\GeV, (ii) the inclusive two-jet ggF process generation, (iii) the \MGvATNLO effective field theory approximation~\cite{Neumann:2016dny,deFlorian:2016spz} normalized to the inclusive N$^{3}$LO cross section.
The $\pt$ spectrum of the Higgs boson for the vector boson fusion (VBF) production mode is re-weighted to account for N$^{3}$LO corrections to the cross section. These corrections~\cite{Cacciari:2015jma,Dreyer:2016oyx} have a negligible effect on the yield for this process for events with Higgs boson $\pt>450\GeV$.

The particle-flow event algorithm~\cite{Sirunyan:2017ulk} is employed to reconstruct and identify each
individual particle with an optimized combination of information from
the various elements of the CMS detector. The algorithm identifies each reconstructed particle as an electron, a muon, a photon, or a charged or a neutral hadron. The missing transverse momentum vector is defined as the negative vectorial sum of the transverse momenta of all the particles identified in the event, and its magnitude is referred to as $\ptmiss$.

The particles are clustered into jets using the anti-$\kt$ algorithm~\cite{Cacciari:2008gp} with a distance parameter of 0.8 (AK8 jets).
To mitigate the effect of pileup, the pileup per particle identification (PUPPI) algorithm~\cite{Bertolini:2014bba} assigns a weight to each particle prior to jet clustering based on the likelihood of the particle originating from the hard scattering vertex. Further corrections are applied to the jet energy as a function of jet $\eta$ and $\pt$ to account for detector response nonlinearities.

To isolate the Higgs boson signal, a high-$\pt$ signal jet is required.
Combinations of several online selections are used, all requiring the total hadronic transverse
energy in the event ($\HT$) or jet $\pt$ to be above a given threshold. In addition, a minimum threshold on the jet mass is imposed after removing remnants of soft radiation with the jet trimming technique~\cite{Krohn:2009th} to reduce the $\HT$ or $\pt$ thresholds and improve the signal acceptance.
The online selection is fully efficient at selecting events offline
with at least one AK8 jet with $\pt > 450\GeV$ and $\abs{\eta} < 2.5$.
Events containing identified and isolated electrons, muons, or $\tau$ leptons with
$\pt > 10$, 10, or 18\GeV and $\abs{\eta} < 2.5$, 2.4, or 2.3, respectively, are vetoed to reduce backgrounds from SM EW processes. Since no genuine $\ptmiss$ is expected for signal processes, events with $\ptmiss >140 \GeV$ are removed in order to further reduce the top quark background contamination.
The leading (in $\pt$) jet in the event is assumed to be the Higgs boson
candidate, the $\PH$ jet.  The soft-drop
algorithm~\cite{Dasgupta:2013ihk,Larkoski:2014wba} is used to remove
soft and wide-angle radiation with a soft radiation fraction $z$
less than 0.1. The parameter $\beta$ is set to zero, which corresponds to the case in which approximately the same fraction of energy is groomed away regardless of the initial jet energy.
The use of soft-drop grooming reduces the jet mass ($\mSD$) for
background QCD events when large jet masses arise from soft gluon
radiation.  For signal events, the jet mass is primarily determined by the
$\PH(\bbbar)$ decay kinematics and its distribution peaks at the mass of the Higgs boson.
Dedicated $\mSD$
corrections~\cite{CMS-PAS-JME-16-003} are derived from simulation and
data in a region enriched with merged
$\PW(\qqbar)$ decays from $\ttbar$ events. They remove a residual dependence on the jet $\pt$
and match the jet mass scale and resolution to those observed in
data.

The dimensionless mass scale variable for QCD jets, $\rho=\mathrm{log}(\mSD^2/\pt^2)$~\cite{Dasgupta:2013ihk,Dolen:2016kst}, whose distribution is roughly invariant in different ranges of jet $\pt$, is used to characterize the correlation between the jet $\cPqb$ tagging discriminator, jet mass, and jet $\pt$.
Only events in the range $-6.0 < \rho < -2.1$ are
considered, to avoid instabilities at the edges of the distribution
due to finite cone limitations from the AK8 jet clustering ($\rho \gtrsim
-2.1$) and to avoid the nonperturbative regime of the soft-drop mass
calculation ($\rho \lesssim -6.0$). This requirement is fully
efficient for the Higgs boson signal.

The $N_2^{1}$ variable~\cite{Moult:2016cvt}, which is based on a ratio of 2-point and 3-point generalized energy correlation functions (ECFs)~\cite{Larkoski:2013eya}, is exploited to determine how consistent a jet is with having a two-prong substructure. The calculation of $N_2^{1}$ is based on the jet constituents after application of the soft-drop grooming algorithm to the jet.  It
provides excellent discrimination between two-prong signal jets and QCD background jets~\cite{Moult:2016cvt}.
However, any selection on $N_2^{1}$ or other similar variables~\cite{Thaler:2010tr} shapes the jet mass distributions differently depending on the $\pt$ of the jet. Therefore a transformation of $N_{2}^{1}$ to $\nddt$ is applied, where DDT stands for designed decorrelated tagger~\cite{Dolen:2016kst}, to reduce its correlation with $\rho$ and $\pt$ in multijet events. We define $\nddt = N_2^{1} - N_2^{1} {}_{(26\%)}$, where $N_2^{1} {}_{(26\%)}$ is the 26th percentile of the $N_2^{1}$ distribution in simulated QCD events as a function of $\rho$ and $\pt$. This ensures that the selection $\nddt<0$ yields a constant QCD background efficiency of 26\% across the entire $\rho$ and $\pt$ range considered in this search. The chosen percentile maximizes the sensitivity to the Higgs boson signal.
In order to select events in which the $\PH$ jet is most likely to contain two $\cPqb$ quarks, we use the double-$\cPqb$ tagger algorithm~\cite{CMS-PAS-BTV-15-002}.
Several observables that characterize the distinct properties of $\cPqb$ hadrons and their flight directions in relation to the jet substructure are used as input variables to this multivariate algorithm, to distinguish between $\PH$ jets and QCD jets. An $\PH$ jet is considered double-$\cPqb$ tagged if its double-$\cPqb$ tag discriminator value is above a threshold corresponding to a 1\% misidentification rate for QCD jets and a 33\% efficiency for $\PH(\bbbar)$ jets.

Events with (without) a double-$\cPqb$ tagged $\PH$ jet define the passing (failing)
region.
In the passing region, the gluon fusion process dominates,
although other Higgs boson production
mechanisms contribute: VBF (12\%), $\PV\PH$ (8\%), $\ttbar\PH$ (5\%). They are all taken into account when extracting the
Higgs boson yield.

The contribution of $\ttbar$ production to the total SM
background is estimated to be less than 3\%. It is obtained from simulation
corrected with scale factors derived from a $\ttbar$-enriched control
sample in which an isolated muon is required.
This sample is included in a global fit used to
extract the signal and the scale factors are treated as unconstrained parameters. They multiply the $\ttbar$ contribution, correcting its overall normalization
and the double-$\cPqb$ mistag efficiency for jets originating from
top quark decays.

The main background in the passing region,
QCD multijet production, has a nontrivial jet mass shape that is difficult to model parametrically and dependent on jet $\pt$
so we constrain it using the signal-depleted failing region.
Since the double-$\cPqb$ tagger discriminator and the jet mass are largely uncorrelated, the passing and failing
regions have similar QCD jet mass distributions, and their ratio,
the ``pass-fail ratio'' $R_{\mathrm{p}/\mathrm{f}}$, is expected to
be nearly constant as a function of jet mass and $\pt$.
To account for the residual difference between the
shapes of passing and failing events, $R_{\mathrm{p}/\mathrm{f}}$ is
parametrized as a polynomial in $\rho$ and $\pt$,
$R_{\mathrm{p}/\mathrm{f}}(\rho,\pt) = \sum_{k, \ell}a_{k\ell} \rho^k \pt^\ell$.
The coefficients $a_{k, \ell}$ have no external constraints but are determined from a simultaneous fit to the data in passing and failing regions across the whole jet mass range.
To determine the order of polynomial necessary to fit the data, a Fisher
$F$-test~\cite{ref:ftest} is performed. Based on its results, a polynomial of second order
in $\rho$ and first order in $\pt$ is selected.

The systematic uncertainties associated with the jet mass scale, the jet
mass resolution, and the $\nddt$ selection efficiency are correlated among the $\PW$, $\PZ$, and $\PH(\bbbar)$ processes.  These
uncertainties are estimated using an independent sample of merged $\PW$ jets. Additional details are available in \suppMaterial.
The efficiency of the double-$\cPqb$ tagger
is measured in data and simulation in a sample enriched in $\bbbar$
from gluon splitting~\cite{CMS-PAS-BTV-15-002}. Scale factors
relating data and simulation are then computed and applied to the
simulation. These scale factors determine the
initial distributions of the jet mass for the $\PW(\qqbar)$, $\PZ(\qqbar)$, and $\PH(\bbbar)$
processes and are further constrained in the fit to data due to the presence of
the $\PW$ and $\PZ$ resonances in the jet mass distribution.
The uncertainty associated with the modeling of the ggF Higgs $\pt$ spectrum is propagated to the overall normalization of the ggF Higgs signal. In addition, the shape of the ggF Higgs $\pt$ distribution is allowed to vary depending on the Higgs boson $\pt$ by up to 30\% at 1000\GeV, without changing the overall normalization.
To account for some potentially $\pt$-dependent deviations due to missing higher-order corrections, uncertainties are applied to the $\PW(\qqbar)$ and $\PZ(\qqbar)$ yields that are $\pt$-dependent and correlated per $\pt$ bin. An additional systematic uncertainty is included to account for potential differences between the $\PW$ and $\PZ$ higher-order corrections (EW $\PW/\PZ$ decorrelation). Finally, additional systematic uncertainties are applied to the $\PW(\qqbar)$, $\PZ(\qqbar)$,
and $\PH(\bbbar)$ yields to account for the uncertainties due to the jet energy scale and
resolution~\cite{jec}, variations in the amount of pileup, and the integrated luminosity
determination~\cite{lumi}. A quantitative summary of the systematic effects considered is shown in Table ~\ref{tab:systematics}.

\begin{table*}
\centering
\topcaption{
Summary of the systematic uncertainties affecting the signal, $\PW$ and $\PZ+$~jets processes. Instances where the uncertainty does not apply are indicated by ``\NA{}".}
\label{tab:systematics}
\begin{scotch}{lcc}
Systematic source & $\PW/\PZ$ & \PH \\
\hline
Integrated luminosity         & 2.5\% &  2.5\% \\
Trigger efficiency & 4\% & 4\%\\
Pileup    &  $<$1\% & $<$1\%\\
$\nddt$ selection efficiency & 4.3\% & 4.3\%\\
Double-$\cPqb$ tag  & 4\% (\PZ)  & 4\% \\
Jet energy scale / resolution  & 10/15\% &  10/15\%\\
Jet mass scale ($\pt$)              &0.4\%/100\GeV ($\pt$) & 0.4\%/100\GeV ($\pt$) \\
Simulation sample size        & 2--25\% & 4--20\% (ggF) \\
$\PH~\pt$ correction  & \NA & 30\% (ggF)\\
NLO QCD corrections &  10\% & \NA\\
NLO EW corrections &  15--35\% & \NA\\
NLO EW $\PW/\PZ$ decorrelation &  5--15\% & \NA\\
\end{scotch}
\end{table*}

In order to validate the background estimation
method and associated systematic uncertainties, studies are performed
on simulated samples injecting signal events and determining the bias on
the measured signal cross section. No significant bias is observed in
these studies.

\begin{figure}[h!]
\centering
    \includegraphics[width=\cmsFigWidth]{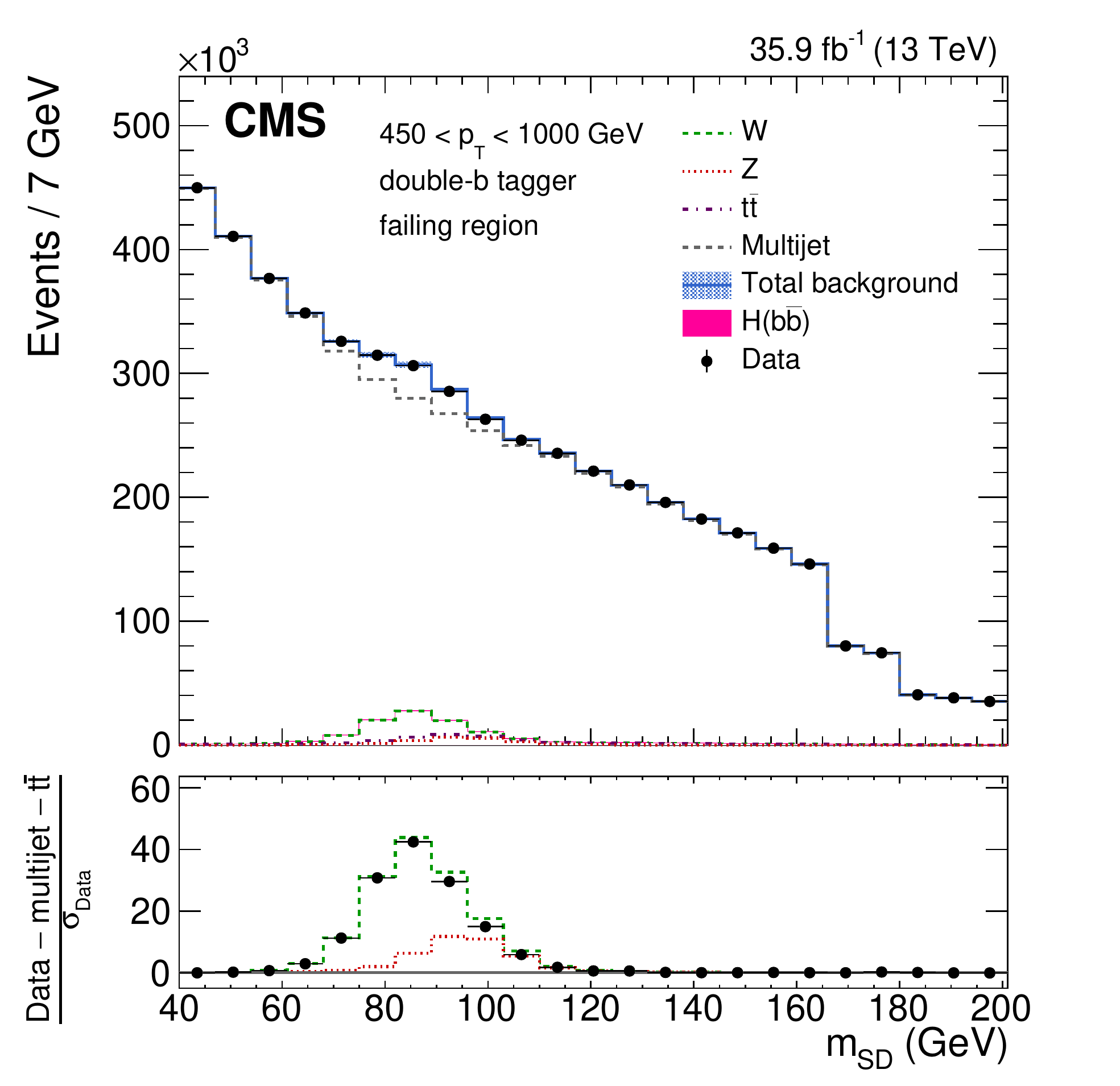}
    \includegraphics[width=\cmsFigWidth]{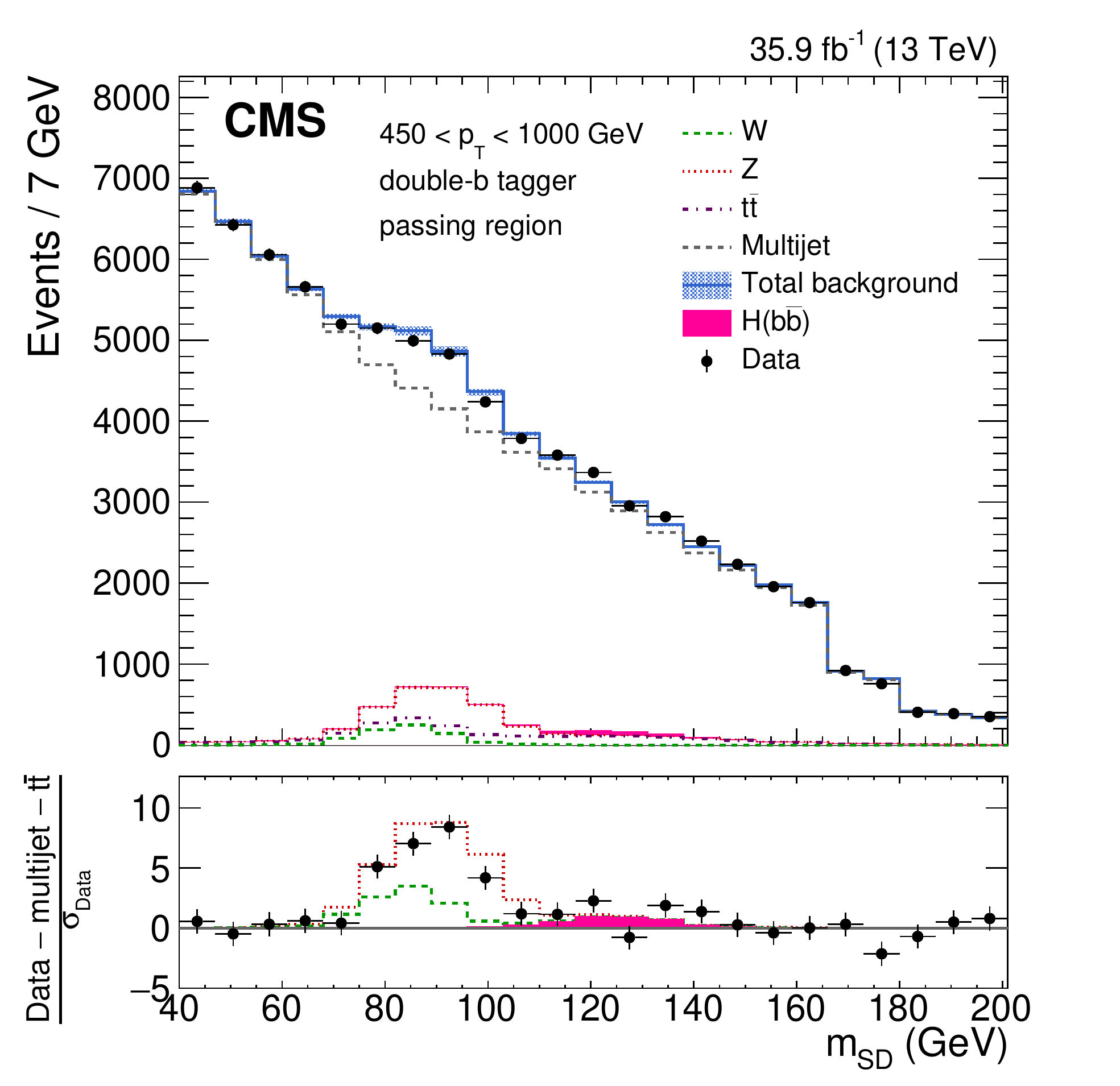}
 \caption{The $\mSD$ distributions in data for the
      failing (\cmsLeft) and passing (\cmsRight) regions and combined $\pt$ categories.
      The QCD multijet background in the passing region is predicted using the failing region
     and the pass-fail ratio $R_{\mathrm{p}/\mathrm{f}}$.
      The features at 166 and 180\GeV in the $\mSD$ distribution are due to the
      kinematic selection on $\rho$, which affects each $\pt$ category differently. In the bottom panel, the ratio of the data to its statistical uncertainty, after subtracting the nonresonant backgrounds, is shown.}
 \label{fig:results}
 \end{figure}

A binned maximum likelihood fit to the observed $\mSD$ distributions in the range 40 to 201\GeV with 7\GeV bin width is performed
using the sum of the $\PH(\bbbar)$, $\PW$, $\PZ$, $\ttbar$, and QCD multijet contributions.
The fit is done simultaneously in the passing and failing regions of the six $\pt$ categories within $450 <\pt <1000\GeV$, and in the $\ttbar$-enriched control region.
The production cross sections relative to the SM cross sections (signal strengths) for the Higgs and the $\PZ$ bosons, $\mu_\PH$ and $\mu_\PZ$, respectively, are extracted from the fit.
Figure~\ref{fig:results} shows the $\mSD$
distributions in data for the passing and failing regions with
measured SM background and $\PH(\bbbar)$ contributions. Contributions
from $\PW$ and $\PZ$ boson production are clearly visible in the data.

The measured $\PZ$ boson signal strength is $\mu_\PZ = \muZVal\pm\muZStatErrLo\stat_{-\muZSystErrLo}^{+\muZSystErrHi}\syst$,  which corresponds to an observed significance of $\muZObsSig$ standard deviations ($\sigma$) with $\muZExpSig\sigma$ expected. This constitutes the first observation of the $\PZ$ boson signal in the single-jet topology~\cite{Aad:1727967} and validates the substructure and $\cPqb$ tagging techniques for the Higgs boson search in the same topology. The measured cross section for the $\PZ+$jets process for jet $\pt>450\GeV$ and $\abs{\eta}<2.5$ is $\xsecZVal\pm\xsecZStatErrLo\stat_{-\xsecZSystErrLo}^{+\xsecZSystErrHi}\syst \unit{pb}$, which is consistent within uncertainties with the SM production cross section of $\xsecZThyVal \pm \xsecZThyErr \unit{pb}$~\cite{Lindert:2017olm}. Likewise, the measured Higgs boson signal strength is $\mu_\PH = \muVal\pm\muStatErrLo\stat_{-\muSystErrLo}^{+\muSystErrHi}\syst$ and includes the corrections to the Higgs boson $\pt$ spectrum described earlier. The corresponding observed (expected) upper limit on the Higgs boson signal strength at a 95\% confidence level is $\muObsLimit$ ($\muExpLimit$), while the observed (expected) significance is $\muObsSig\sigma$ ($\muExpSig\sigma$).
The observed $\mu_\PH$ implies a measured ggF cross section times $\PH(\bbbar)$ branching fraction for jet $\pt>450\GeV$ and $\abs{\eta}<2.5$ of $\xsecVal\pm\xsecStatErrLo\stat_{-\xsecSystErrLo}^{+\xsecSystErrHi}\syst\unit{fb}$, assuming the SM values for the ratios of the different $\PH(\bbbar)$ production modes. This measurement is consistent within uncertainties with the SM ggF cross section times $\PH(\bbbar)$ branching fraction of $\xsecThyVal\pm\xsecThyErr\unit{fb}$.

\begin{figure}[h!]
\centering
\includegraphics[width=\cmsFigWidth]{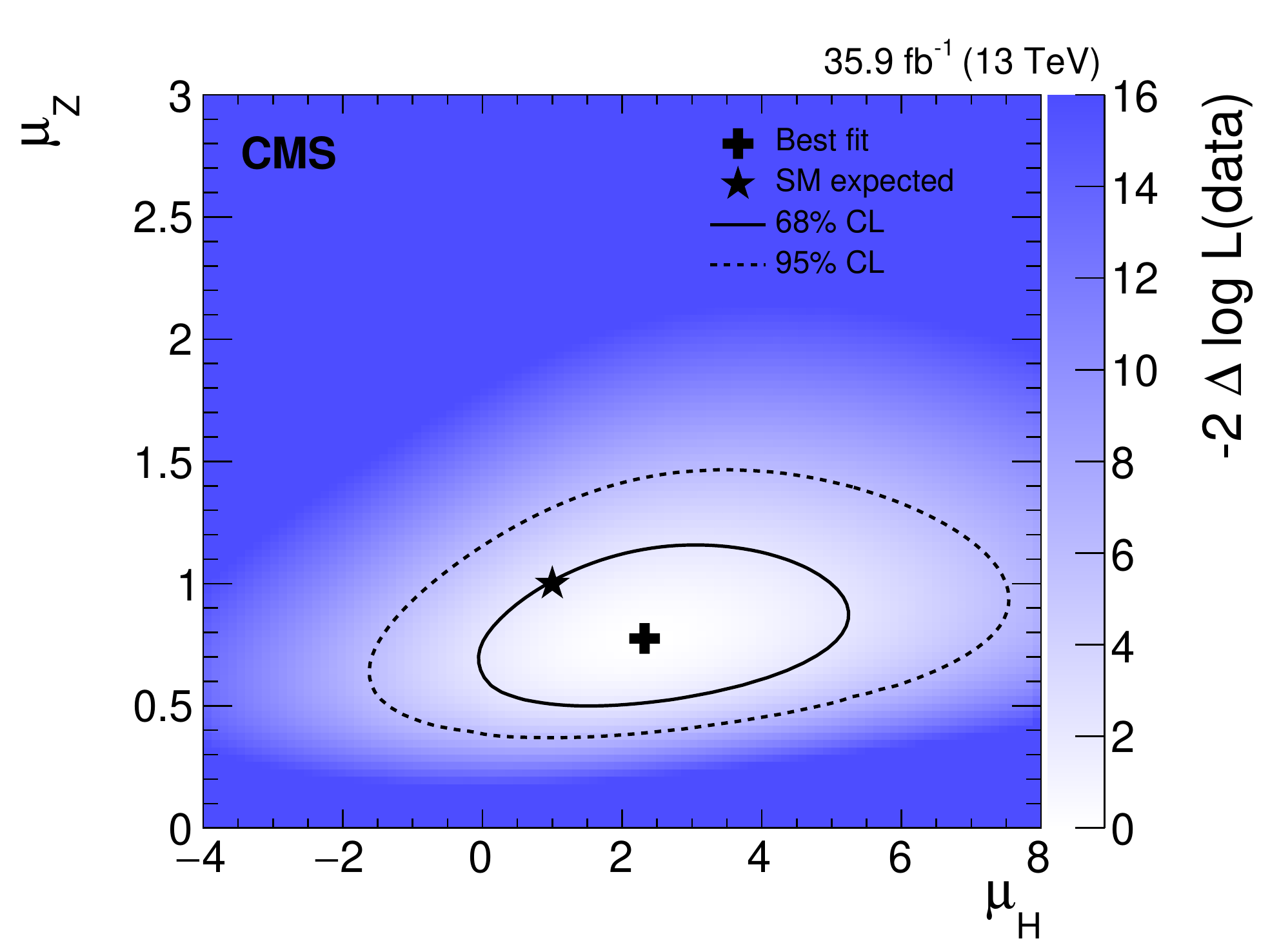}
        \caption{Profile likelihood test statistic $-2\Delta\log\mathcal{L}$ scan in data as a function of the Higgs and
          $\PZ$ bosons signal strengths ($\mu_\PH, \mu_\PZ$).}
        \label{fig:deltaLLZ}
 \end{figure}

Table ~\ref{tab:ObservedSig} summarizes the measured signal strengths and significances for the Higgs and $\PZ$ boson processes. In particular, they are also reported for the case in which no corrections to the Higgs boson $\pt$ spectrum are applied.
Figure~\ref{fig:deltaLLZ} shows the profile likelihood test statistic scan in data as function of the Higgs and $\PZ$ boson signal strengths ($\mu_{\PH},\mu_{\PZ}$).

\begin{table}[h!]
\centering
 \topcaption{Fitted signal strength, expected and observed significance of the Higgs and $\PZ$ boson signal. The 95\% confidence level upper limit (UL) on the Higgs boson signal strength is also listed. }
\begin{scotch}{ llll }
 & $\PH$ & $\PH$ no \pt corr. & $\PZ$ \\
Observed signal strength & $\muVal_{-\muErrLo}^{+\muErrHi}$ & $\muNoVal_{-\muNoErrLo}^{+\muNoErrHi}$  & $\muZVal_{-\muZErrLo}^{+\muZErrHi}$\\
Expected UL signal strength   & $<\muExpLimit$  & $<\muNoExpLimit$ & --- \\
Observed UL signal strength   & $<\muObsLimit$  & $<\muNoObsLimit$ & --- \\
Expected significance & $\muExpSig\sigma$ & $\muNoExpSig\sigma$ &  $\muZExpSig\sigma$ \\
Observed significance & $\muObsSig\sigma$ & $\muNoObsSig\sigma$ &  $\muZObsSig\sigma$\\
\end{scotch}
\label{tab:ObservedSig}
\end{table}

In summary, an inclusive search for the standard model Higgs boson with $\pt > 450 \GeV$ decaying to bottom quark-antiquark pairs and reconstructed as a single, large-radius jet is presented.
The $\PZ+$jets process is observed for the
first time in the single-jet topology with a significance of
$\muZObsSig\sigma$. The Higgs production is measured
with an observed (expected) significance of $\muObsSig\sigma$
($\muExpSig\sigma$) when including Higgs boson $\pt$ spectrum corrections accounting for higher-order and finite top quark mass effects. The measured cross section times branching fraction for the gluon fusion $\PH(\bbbar)$ production for reconstructed $\pt>450\GeV$ and $\abs{\eta}<2.5$ is
$\xsecVal\pm\xsecStatErrLo\stat_{-\xsecSystErrLo}^{+\xsecSystErrHi}\syst\unit{fb}$,
which is consistent with the SM prediction within uncertainties.

\begin{acknowledgments}
We congratulate our colleagues in the CERN accelerator departments for the excellent performance of the LHC and thank the technical and administrative staffs at CERN and at other CMS institutes for their contributions to the success of the CMS effort. In addition, we gratefully acknowledge the computing centers and personnel of the Worldwide LHC Computing Grid for delivering so effectively the computing infrastructure essential to our analyses. Finally, we acknowledge the enduring support for the construction and operation of the LHC and the CMS detector provided by the following funding agencies: BMWFW and FWF (Austria); FNRS and FWO (Belgium); CNPq, CAPES, FAPERJ, and FAPESP (Brazil); MES (Bulgaria); CERN; CAS, MoST, and NSFC (China); COLCIENCIAS (Colombia); MSES and CSF (Croatia); RPF (Cyprus); SENESCYT (Ecuador); MoER, ERC IUT, and ERDF (Estonia); Academy of Finland, MEC, and HIP (Finland); CEA and CNRS/IN2P3 (France); BMBF, DFG, and HGF (Germany); GSRT (Greece); OTKA and NIH (Hungary); DAE and DST (India); IPM (Iran); SFI (Ireland); INFN (Italy); MSIP and NRF (Republic of Korea); LAS (Lithuania); MOE and UM (Malaysia); BUAP, CINVESTAV, CONACYT, LNS, SEP, and UASLP-FAI (Mexico); MBIE (New Zealand); PAEC (Pakistan); MSHE and NSC (Poland); FCT (Portugal); JINR (Dubna); MON, RosAtom, RAS, RFBR and RAEP (Russia); MESTD (Serbia); SEIDI, CPAN, PCTI and FEDER (Spain); Swiss Funding Agencies (Switzerland); MST (Taipei); ThEPCenter, IPST, STAR, and NSTDA (Thailand); TUBITAK and TAEK (Turkey); NASU and SFFR (Ukraine); STFC~(United Kingdom);~DOE and NSF~(USA).
\end{acknowledgments}
\bibliography{auto_generated}
\ifthenelse{\boolean{cms@external}}{}{
\clearpage
\appendix
\section{Additional material on systematic uncertainties associated with the jet mass scale, the jet mass resolution, and \texorpdfstring{$N_2^\text{1,DDT}$}{N2(1,DDT)} selection\label{app:suppMat}}
The systematic effects on the shapes and normalization of the \PW, \PZ{} backgrounds and signal components are strongly correlated since they are affected by similar systematic mismeasurements. These uncertainties are estimated using an independent sample of merged \PW~boson jets in semileptonic \ttbar events in data. In this region, events are required to have an energetic muon with $\pt > 100$\GeV, $\pt^\text{miss} > 80$\GeV,
a high-\pt AK8 jet with $\pt > 200$\GeV, and an additional jet separated from the AK8 jet by $\Delta R > 0.8$. Using the same $N_2^\text{1,DDT}$ requirement described in the text, we define samples with events that pass and fail the selection for merged \PW~boson jets in data and simulation. A simultaneous fit to the two samples is performed in order to extract the selection efficiency of a merged \PW~jet in simulation and in data.
We measure the data-to-simulation scale factor for the $N_2^\text{1,DDT}$ selection to be $0.99 \pm 0.04$.  The mass scales in data and simulation are found to be consistent within 1\%. The jet mass resolution data-to-simulation scale factor is
measured to be $ 1.08 \pm 0.09 $.
These scale factors are applied to the merged \PZ{} and \PH{} jets. Systematic uncertainties related to the differences between hadronic decays of \PW{} and \PZ{} or \PH{} bosons are estimated by comparing the \PYTHIA and \HERWIGpp~2.7.1 \cite{Bahr:2008pv} showering algorithms, and found to be negligible.
They determine the initial distributions of the jet mass for the $\PW(\qqbar)$, $\Z(\qqbar)$, and $\PH(\bbbar)$ processes and are further constrained in the fit to data due to the presence of the W and Z resonances in the jet mass distribution. This control sample does not provide a measurement for jets with very high \pt. An additional systematic uncertainty is included to account for the extrapolation of this measurement to very high \pt jets. The jet mass scale uncertainty is allowed to vary in the signal extraction differently depending on the jet \pt (from 2\% at 500\GeV to 4\% at 1\TeV). The simultaneous extraction of the \PZ{} and \PH{} boson signal strengths aims to calibrate the relative jet mass scale in situ.  This is the first measurement of this kind at such high \pt. In the fit to data, the jet mass scale is constrained well within the uncertainty we quote.

}

\cleardoublepage \section{The CMS Collaboration \label{app:collab}}\begin{sloppypar}\hyphenpenalty=5000\widowpenalty=500\clubpenalty=5000\textbf{Yerevan Physics Institute,  Yerevan,  Armenia}\\*[0pt]
A.M.~Sirunyan, A.~Tumasyan
\vskip\cmsinstskip
\textbf{Institut f\"{u}r Hochenergiephysik,  Wien,  Austria}\\*[0pt]
W.~Adam, F.~Ambrogi, E.~Asilar, T.~Bergauer, J.~Brandstetter, E.~Brondolin, M.~Dragicevic, J.~Er\"{o}, M.~Flechl, M.~Friedl, R.~Fr\"{u}hwirth\cmsAuthorMark{1}, V.M.~Ghete, J.~Grossmann, J.~Hrubec, M.~Jeitler\cmsAuthorMark{1}, A.~K\"{o}nig, N.~Krammer, I.~Kr\"{a}tschmer, D.~Liko, T.~Madlener, I.~Mikulec, E.~Pree, N.~Rad, H.~Rohringer, J.~Schieck\cmsAuthorMark{1}, R.~Sch\"{o}fbeck, M.~Spanring, D.~Spitzbart, W.~Waltenberger, J.~Wittmann, C.-E.~Wulz\cmsAuthorMark{1}, M.~Zarucki
% ...
\vskip\cmsinstskip
% ...
1:~~Also at Vienna University of Technology, Vienna, Austria\\
% ...

\end{sloppypar}
\end{document}